# Switching from pyroelectric to ferroelectric order in Ni doped $CaBaCo_4O_7$

C.Dhanasekhar[1,5], A. K Das[1], Ripandeep Singh[2], A. Das[2], G. Giovannetti[3], D. Khomskii[4], A.Venimadhav[5*]

[1,5]Department of physics, Indian Institute of Technology, Kharagpur -721302, India

[2]Solid State Physics Division, Bhabha Atomic Research Centre, Mumbai 400085, India

[3]Consiglio Nazionale delle Ricerche-Istituto Nazionale per la Fisica della Materia (CNR-INFM), CASTI Regional Laboratory, 67100 L'Aquila, Italy

[4]Physikalisches Institut, Universitätzu Köln, ZülpicherStrasse 77, 50937 Köln, Germany

[5]Cryogenic Engineering Centre, Indian Institute of Technology, Kharagpur -721302, India

**Abstract**

We report ferroelectric ordering in Ni substituted $CaBaCo_4O_7$. Magnetization showed ferrimagnetic transition at 60 K and an additional transition is found ~ 82 K, further, enhanced antiferromagnetic interactions and decrease in saturation magnetization are noticed with Ni substitution. The dielectric and pyroelectric measurements illustrate a strong coupling between spin and charge degrees of freedom; ferroelectric behavior is confirmed with enhanced ordering temperature (~82 K) and saturation polarization (250 $\mu C/m^2$). Neutron diffraction has revealed an increase in c-lattice parameter in Ni sample and all the Co/Ni moments are reoriented in a- direction; evidently a non-collinear ferrimagnetic to collinear ferrimagnetic spin order is observed. The coupling between the triangular and Kagome layers weakens and leads to ↑↑↓↓ AFM ordering in the Kagoma layer. This can be viewed as a 2D-collinear layer with unequal bond distances and most likely responsible for the switching of electric polarization.

Multiferroics having simultaneous presence of magnetic and electric dipolar ordering having the ability to couple electric and magnetic polarization are potential materials for important novel applications. Though simultaneous presence of ferromagnetism and ferroelectricity is rather uncommon, since the pioneering works on $BiFeO_3$ and $TbMnO_3$ a number of magnetically induced ferroelectric materials were discovered [1-4]. In such systems, the symmetry breaking necessary for the dipolar ordering is facilitated by non-collinear spin ordering through inverse Dzyaloshinskii-Moriya (DM) or spin orbit coupling [5-7]. Alternately, collinear spin structure has shown promising magnetoelectric coupling through symmetric exchange striction mechanisms in $RMnO_3$, $RMn_2O_5$ and Ising spin chain $Ca_3CoMnO_6$ [8-10]. In recent years the *Swedenborgite* (SB) minerals having general formula of $ABaCo_4O_7$ (A= Rare earth, Ca) have gained large attention due to their polar crystal structure [11-14]. More interestingly, they contain geometrically frustrated alternating triangular (T) and kagome (K) layers with Co tetrahedra, with Co ions in mixed valence states of 2+ and 3+. The ratios of these valences can be changed from 3: 1 to 1:1 by replacing the rare earth by calcium. In $CaBaCo_4O_7$ (CBCO), an orthorhombic structure (with *Pbn2$_1$*) is realized due to strong buckling of the $CoO_4$ tetrahedra in the kagome layers which partially releases the frustration. In contrast to RBaCo4O7, in CBCO a charge ordering of $Co^{2+}$ and $Co^{3+}$ is realized [14, 15]. A ferrimagnetic ordering with net moment along b-direction is suggested from the neutron diffraction studies [15].

In CBCO, an exceptionally large electric polarization (17000 $\mu C/m^2$) is found along the polar axis, below the ferrimagnetic ordering. However, due to non-switchable electric polarization this material is classified not as ferroelectric (FE), but as pyroelectric. Johnson *et al.* have theoretically confirmed the pyroelectric nature of CBCO in both magnetically ordered and disordered phases [12]. On the other hand Fishman *et al.* have presented a slightly different magnetic structure for the CBCO by combining the neutron diffraction, magnetization with terahertz measurements, and the giant electric polarization is mainly attributed to the set of bond in the bitetrahedral Co chains along c-axis [14].

The distinction of pyroelectricty and ferroelectricity is significant for many applications, because the energy barrier between the two oppositely polarized states is infinite for pyroelectric materials, but for most applications the possibility of switching of electric polarization is crucial. In order to take advantage of multiferroicity, ferroelectric ground state is desirable. Below we report the transformation of pyroelectric to ferroelectric state with enhanced polarization and enhanced ordering temperature in Ni-substituted CBCO. A ferrimagnetic ordering with resulting moment in a- direction is found instead of b-direction as in the case of CBCO. Here, the kagome layers have antiferromagnetic ordering with ↑↑↓↓ spin structure as revealed by neutron diffraction study. The unequal bond distances of ↑↑ and ↑↓ neighbours suggest magnetoelectric coupling through magnetostriction [26].

The $CaBaCo_{4-x}Ni_xO_7$ (x=0 (CBCO) and 0.10 (CBCNO)) samples were prepared by solid state reaction method [16, 22,] and the details of synthesis and characterization is given in the supplementary material (SM) [23]. Temperature dependence of ac magnetization for CBCO and CBCNO samples is shown in Fig.1 (a and b). Both CBCO and CBCNO sample show the ferrimagnetic ordering at 60 K, while CBCNO shows an additional transition at 82 K. In contrast to the CBCO sample, the ac magnetization of the CBCNO increases below 40 K. Temperature dependence of dc magnetization for CBCNO samples also shows two magnetic transitions at 60 K and 82 K as shown in Fig.1 (b). Further, dc magnetization increase below 40 K, consistent with the ac magnetization behavior. Though there is no clear understanding of the 82 K transition, its presence was reported in CBCO with $Sr^{2+}$ doping at Ca site and $Zn^{2+}$, $Fe^{3+}$ doping at Co site [17-19].

M-H measured at 5 and 45 K for the CBCO and CBCNO samples is shown in the Fig.1 (c-f). CBCNO shows the decrease of coercive field and saturation magnetic moment at 5 K. More importantly, butterfly shape M-H loops is noticed in CBCNO, In fact this has been notices below 60 K and it hints the increased AFM interaction in CBNCO. The results suggest that the magnetic behaviour of CBNCO is more complex than that of CBCO.

The temperature dependent dielectric permittivity of the CBCO and CBCNO samples is shown in the Fig. 2 (a) and (c). CBCO shows a peak at 60 K that matches with the magnetic ordering temperature. The CBCNO samples show two peaks at 60 K and 82 K, which matches with the magnetic transitions. Interestingly, the dielectric permittivity is higher in CBCNO.

Interestingly, varying natures of electric polarization is reported in CBCO, from the reversal of ΔP in bulk sample (ferroelectric) to a non-switchable ΔP (pyroelctric) in single crystals along polar axis (c-axis) and along b axis [11,20-21]. Hence, it is necessary to reveal the dipolar ground state of our bulk CBCO. Pyrocurrent ($I_p$) measurement can distinguish between pyroelectric and ferroelectric samples depending on the poling field ($E_p$). Fig. 2 (b) summarizes the pyroelectric behaviour of CBCO sample; only positive pyrocurrent + $I_p$ is observed for both positive and negative poling fields ± $E_p$ (=150 kV/m); corresponding + ΔP is shown in Fig. 2(b). Further, to confirm the pyroelectric behaviour, we have measured the $I$p in heating and cooling paths in absence of $E_p$, and the results are shown in the inset of Fig. 2(b). The sign of the $I_p$ is opposite for cooling and heating paths, and this behaviour is typical for pyroelectric materials (see the SM Fig.S1 [23]) [22].

The inset of Fig.2 (d) shows the comparison of $I_p$ peaks for the CBCO and CBCNO samples measured under a poling field of +150 kV/m. The $I_p$ peak of CBCNO, becomes broader compared to CBCO and correlates with the magnetic and dielectric transitions. The most important result is presented in Fig. 2(d), where theΔP behaviour of CBCNO sample for $E_p$ of ±150 kV/m is shown. The switching of the ΔP for the ± $E_p$ suggests the ferroelectric nature of the CBCNO sample. The polarization magnitude of 100 μC/m$^2$ is higher than that of the CBCO sample; also the absence of the shift in temperature of the $I_p$ peak (Fig. 2 (h)) with different heating rates supports the ferroelectric nature of CBCNO. At low temperatures, the ΔP increases up to $E_p$ ~ 300 kV/m (ΔP =250 μC/m$^2$) without saturation, which suggests the need for higher poling field to saturate the electric dipoles.

In order to verify the intrinsic nature of ferroelectric behaviour we have conducted additional $I_p$ measurements to obtain spontaneous ferroelectric polarization due to permanent dipoles [24, 25]. Here the $E_p$ is applied only once above the transition temperature and cooled down to low temperature (~10 K). Then the sample short circuited and subjected to long rest ($h_{sc}$ 30 mins), later conducted subsequent cooling and heating cycles to obtain the spontaneous polarization. Initially the sample was cooled under $E_p$ (+300 kV/m) from 85-10 K. At 10 K the $E_p$ was removed and the sample was shorted for 30 minutes. Now, without $E_p$, the sample was heated (Fig.2 (f); path 1) with a ramp rate of 5 K/min up to 50 K (< the maximum $I_p$ peak temperature of 65 K) and the heating was stopped for a halt time ($h_t$) of 5 mins until the $I_p$ vanishes (path 1). After that the sample is cooled from 50 K-10 K (path 2); the $I_p$ goes to negative and recovers to a stable current at low temperature as shown in Fig. 2 (f). The negative $I_p$ is expected as dT/dt is negative. In the last step (Fig.2 (f); path-3), sample is again heated from 10 K to 90 K with the same rate and obtained a peak at ~ 70 K. The experiment was repeated for another $h_t$ of 30 minutes. The magnitude of $I_p$ obtained is same for both the halt times confirms the spontaneous nature of polarization (inset of Fig. 2(f)). Further, we have also confirmed the switching of the $I_p$ by varying the temperature sinusoidally in between 40-42 K as shown in the Fig. 2 (g). Note that the temperature cycling was done after waiting at 50 K for 30 minutes (after path 1). The switching of $I_p$ and phase shift of $90^0$ between $I_p$ and temperature is reminiscent to ferroelectric samples. The above experiments exclude extrinsic character of the observed ferroelectric behavior, due e.g. thermally stimulated free charge carriers and charge accumulation at the grain boundaries. Ferroelectric behaviour is not only found in CBCNO with 10% of Ni, but also found in samples with lower Ni (5 %) substitutions (see the SM Fig.S2 [23]).

Thus, the magnetic and transport measurements suggest two important differences of CBCNO from that of the parent sample: (a) CBCNO is ferrimagnetic with decreased magnetic moment and with dominant antiferromagnetic interactions. (b) CBCNO demonstrates the ferroelectric behavior instead of the pyroelectric one in CBCO. Neutron diffraction can shed some light on the origin of these changes.

The low temperature neutron powder diffraction (NPD) data of the CBCO and CBCNO samples are shown in Fig. 3(a) & (b). The structural refinement on both the samples confirms the orthorhombic *Pbn2$_1$* symmetry. A minor impurity phase of CoO [~ 3 wt %] is identified in both the samples. The detailed structural parameters and average bong lengths for the CBCO and CBCNO samples at 300 K and 6 K are given in SM [23].The lattice parameters obtained at 6 K for CBCO are a= 6.2556(5) Å, b =11.0383(9) Å, c= 10.1563(8) Å and for CBCNO they are a= 6.2561(6) Å, b=11.0246(9) Å and c= 10.1673(9) Å. CBCNO shows reduction in volume (see TABLE.II. SM [23]), consistent with the smaller ionic radii of $Ni^{2+}$ ion. And Ni occupied in charge ordered state of CBCO by $Co^{2+}$, as expected due to a stable valence of $Ni^{2+}$. The lattice parameters 'b' found to be smaller in CBCNO, while the 'c'-parameter has increased. Although the magnetic structure of the CBCO and CBCNO is represented with similar irreducible representation, the intensities of magnetic reflections are different in both samples (shown in right side of Figs.3 (a) and (b)). The temperature variation of the (012) magnetic reflection [shown in inset of Fig.3 (b)] of the CBCNO confirms the ferrimagnetic transition at ~ 60 K; whereas in case of CBCO, (101) magnetic reflection governs the ferrimagnetic ordering at this temperature.

The spin structures of CBCO and CBCNO are shown in Fig.3(c) and (d). The magnetic structure for the CBCO is in agreement with Caignaert *et al*. (shown in Fig.3(c)) where "Co2 Co3" zigzag ferromagnetic chains running along b direction couple antiferromagnetically with Co4 and Co1 (in T layer) spins [15]. The resultant ferrimagnetic moment is in b-direction for CBCO. Recently, Fishman *et al.* pointed out that the easy-plane and easy-axis anisotropies within each K and T layers plays a major role to induce the non-collinear ferrimagnetic ordering in CBCO [14]. In CBNCO, an increase in c-parameter compared to the CBCO suggests the weakening of interactions between the adjacent T and K-layers: this leads to the spin reorientation and the collinear magnetic structure. The Co (Ni) 2 -

Co (Ni) 3 and Co (Ni) 2 –Co4 spins in the kagome layer form ↑↑↓↓ AFM spin order in the ab plane as shown in the Fig 3 (d). And Co1 spins of T-layer are antiferromagnetically coupled with the ↑↑↓↓ spin structure (formed by Co (Ni) 2- Co (Ni) 3–Co4) in the K-layer) results in an overall collinear ferrimagnetic ordering with the net moment in a-direction. The overall decrease of saturation moment with Ni supports the neutron spin structure [Table. IIa. SM [23]].The 82 K transition in CBCNO samples may not be associated with the long range ordering, since the main magnetic reflection (0 1 2) disappears above 65 K. We suppose that the short range correlations arising from the K-layers can be responsible for this magnetic anomaly.

What could be the reasons for the change from the pyroelectric behavior of the undoped CBCO to ferroelectric one observed in CBCNO? One natural candidate could be the change from noncollinear to a collinear ↑↑↓↓ magnetic structure discussed above. It is well documented that such ordering can naturally produce (switchable) ferroelectric polarization, as is observed in E-type manganites [8] or in $Ca_3CoMnO_6$ [10], and as is theoretically discussed e.g. in [26, 27]. The same mechanism (formation of inequivalent, short and long ↑↓ and ↑↑ bonds or respective shifts of oxygens, could also be efficient in this case (see Fig.4). A switchable polarization in each kagome layer would then point in b-direction, not along the original pyroelectric c-axis. However, detailed analysis shows that in the simple form this mechanism may not work here: polarization in the neighbouring kagome layers seem to be opposite and compensate each other. Thus in this model CBCNO would be not ferroelectric but rather antiferroelectric.

Our *ab intio* calculations confirm this conclusion. The electric polarization was calculation using the Berry phase method [28]. The outcome of our calculations is that indeed there appears some extra polarization in magnetically-ordered state, but it is directed along c, as in undoped case [12], however, the polarizations in b-direction, discussed above, apparently cancel in neighbouring kagome layers. Thus this simple picture in its pure form does not explain our experimental observations. One can argue that if by some reasons kagome layers would become inequivalent, e.g. if there is a tendency of Ni ions to segregate in every second layer, this compensation would not work and the material would indeed become ferroelectric. At the moment we do not have experimental indications that it might be the case. But in any case, as discussed above, we believe that the FE behavior of CBCNO is intrinsic, though its detailed microscopic explanation is still missing.

In summary, Ni substitution in $CaBaCo_4O_7$ (CBCO) produces ferroelectricity necessary for spintronic applications. The dielectric, magnetic and pyrocurrent measurements shows strong correlation between electric and magnetic degrees of freedom in CBCNO samples. Ni substitution drastically modifies the spin structure from non-collinear to a collinear ferrimagnetic order. Collinear spin structure-driven multiferroics are not many, and CBCNO seems particularly interesting having 2D ↑↑↓↓ structure in the kagome layers. The exchange striction in the ↑↑↓↓ collinear structure could in principle be the driving force for the ferroelectric behavior, although at the moment we do not have full explanation of the observed behavior. The 2D nature of the magnetic structure seems interesting for designing new magnetoeletric materials and can be extended to artificial epitaxial thin films.

**Acknowledgements**

The authors of IIT Kharagpur acknowledge DST, India for FIST project and IIT Kharagpur funded VSM SQUID magnetometer. The work of D.Kh. was supported by the German project SFB 1238 and by the Koeln Univertsity via German Excellence initiative.

[*] venimadhav@hijli.iitkgp.ernet.in

**REFERENCES**

[1] Daniel Khomskii, Physics **2**, 20 (2009).

[2] R. Ramesh & Nicola A. Spaldin, Nat. Mater. **6**, 21 (2007).

[3] Sang-Wook Cheong & Maxim Mostovoy, Nat. Mater. **6**, 13 (2007).

[4] Yoshinori Tokura, Shinichiro Seki and Naoto Nagaosa, Rep. Prog. Phys. **77**, 076501 (2014).

[5] Hosho Katsura, Naoto Nagaosa, and Alexander V. Balatsky, Phys. Rev. Lett. **95**, 057205 (2005).

[6] I. A. Sergienko and E. Dagotto, Phys. Rev. B **73**, 094434 (2006).

[7] H. Murakawa, Y. Onose, S. Miyahara, N. Furukawa, and Y. Tokura, Phys. Rev. Lett. **105**, 137202 (2010).

[8] Ivan A. Sergienko, Cengiz Şen, and Elbio Dagotto, Phys. Rev. Lett. **97**, 227204 (2006).

[9] Hua Wu, T. Burnus, Z. Hu, C. Martin, A. Maignan, J. C. Cezar, A. Tanaka, N. B. Brookes, D. I. Khomskii, and L. H. Tjeng, Phys. Rev. Lett. **102**, 026404 (2009).

[10] Y. J. Choi, H. T. Yi, S. Lee, Q. Huang, V. Kiryukhin, and S.-W. Cheong, Phys. Rev. Lett. **100**, 047601 (2008).

[11] V. Caignaert, A. Maignan, K. Singh, Ch. Simon, V. Pralong, B. Raveau, J. F. Mitchell, H. Zheng, A. Huq, and L. C. Chapon, Phys. Rev. B **88**, 174403 (2013).

[12] R. D. Johnson, K. Cao, F. Giustino, and P. G. Radaelli, Phys. Rev. B **90**, 045129 (2014).

[13] S. Bordács, V. Kocsis, Y. Tokunaga, U. Nagel, T. Rõõm, Y. Takahashi, Y. Taguchi, and Y. Tokura, Phys. Rev. B **92**, 214441 (2015).

[14] R. S. Fishman, S. Bordács, V. Kocsis, I. Kézsmárki, J. Viirok, U. Nagel, T. Rõõm, A. Puri, U. Zeitler, Y. Tokunaga, Y. Taguchi, and Y. Tokura, Phys. Rev. B **95**, 024423 (2017).

[15] V. Caignaert, V. Pralong, V. Hardy, C. Ritter, and B. Raveau, Phys. Rev. B **81**, 094417 (2010).

[16] C. Dhanasekhar, A.K. Das, A. Venimadhav, J. Magn. Magn. Mater. **418**, 76–80 (2016).

[17] G. Aurelio, J. Curiale, F. Bardelli, R. Junqueira Prado, L. Hennet, G. Cuello, J. Campo, and D. Thiaudière, J. Appl. Phys. **118**, 134101 (2015).

[18] Tapati Sarkar, Md. Motin Seikh, V. Pralong, V. Caignaert and B. Raveau, J. Mater. Chem., **22**, 18043 (2012). Md. Motin Seikh, V. Caignaert, E. Suard, K. R. S. Preethi Meher, A. Maignan, and B. Raveau, J. Appl. Phys. **116**, 244106 (2014).

[19] Md. Motin Seikh, Asish K. Kundu, V. Caignaert, B. Raveau, J. Alloys Compd. **656**, 166-171 (2016).

[20] K. Singh, V. Caignaert, L. C. Chapon, V. Pralong, B. Raveau, and A. Maignan, Phys. Rev. B **86**, 024410 (2012).

[21] H. Iwamoto, M. Ehara, M. Akaki, and H. Kuwahara, J. Phys: Conference Series **400**, 032031 (2012).

[22] Sidney B. Lang, Phys. Today **58**, 31–36 (2005).

[23] See Supplemental Material at http://link.aps.org/ supplemental/****/Phys.Rev.B.**** for details of sample preparation, structural parameters obtained from neutron diffraction and polarization studies on 5% Ni doped sample.

[24] Hariharan Nhalil, Harikrishnan S. Nair, C. M. N. Kumar, André M. Strydom, and Suja Elizabeth, Phys. Rev. B **92**, 214426 (2015).

[25] Lynn E. Garn and Edward J. Sharp, J. Appl. Phys. **53**, 8974 (1982). Edward J. Sharp and Lynn E. Garn, J. Appl. Phys. **53**, 8980 (1982).

[26] Jeroen van den Brink, and Daniel I Khomskii, J. Phys. Condens. Matter **20**, 434217 (2008).

[27] Gianluca Giovannetti, Sanjeev Kumar, Daniel Khomskii, Silvia Picozzi, and Jeroen van den Brink, Phys. Rev. Lett. **103**, 156401 (2009).

[28] R. D. King-Smith and David Vanderbilt, Phys. Rev. B **47**, 1651 (1993). Raffaele Resta, Rev. Mod. Phys. **66**, 899 (1994).

**Figures**

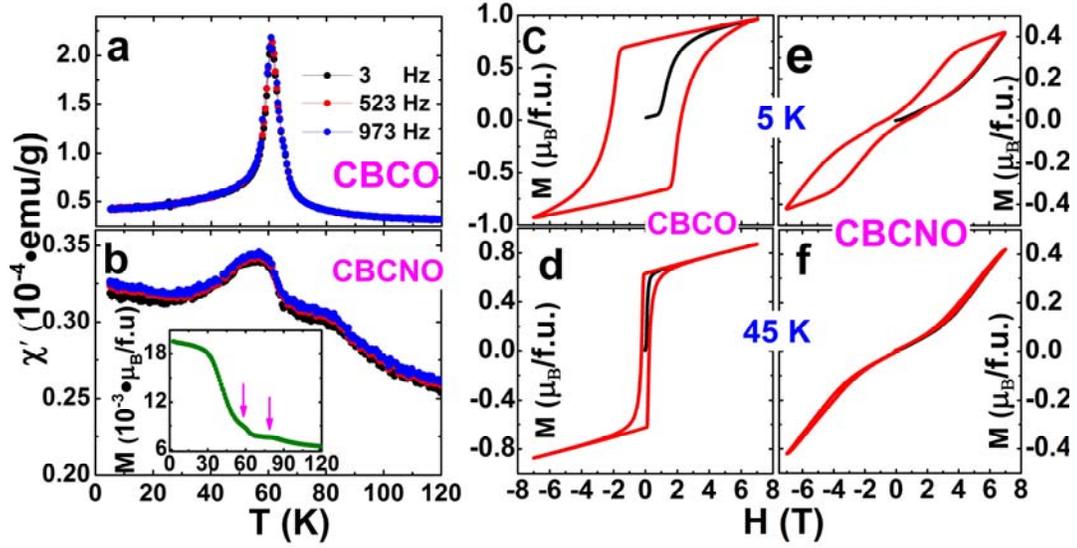

**FIG.1.** (Color online). (a) and (b) Real part of ac magnetization vs temperature for CBCO and CBCNO samples measured under $H_{ac}=1Oe$; Inset of (b) shows the field-cooled dc magnetization for CBCNO sample measured under 3000 Oe. (c-f) show the M vs H measured at various temperatures for CBCO and CBCNO samples.

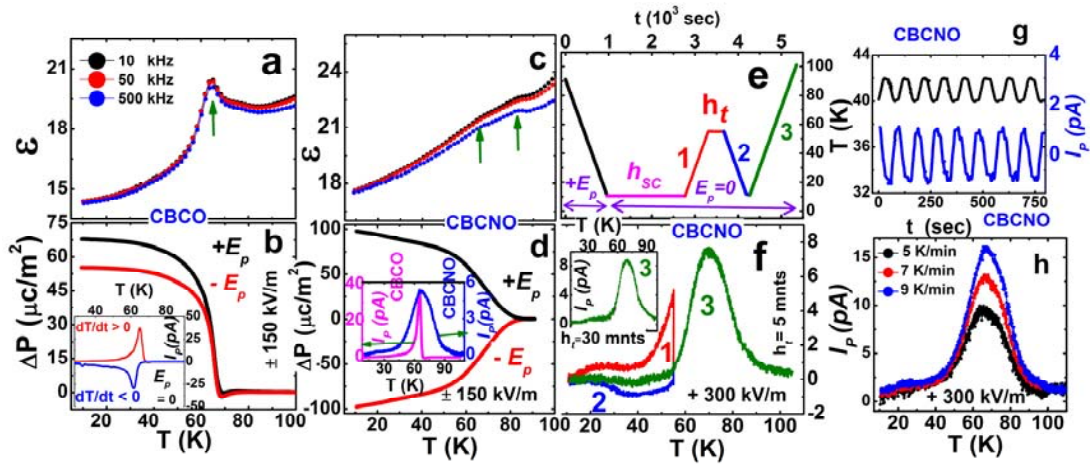

**FIG.2.** (Color online). (a) Temperature dependent dielectric constant for CBCO and (c) CBNCO samples; (b) $\Delta P(T)$ of CBCO; inset of (b) shows $I_p(T)$ of CBCO measured under zero $E_p$. (d) $\Delta P(T)$ of CBCNO; (e) time vs. temperature of the measurement protocol; (f) $I_p(T)$ for CBCNO measured using protocol described in text. Inset of (f) shows $I_p(T)$ for 30 mins halting at 50 K. (h) heating rate dependence of $I_p(T)$ *measured* under $E_p$ ±300 kV/m. (g) $I_p$ versus temperature cycling after 30 minutes of waiting.

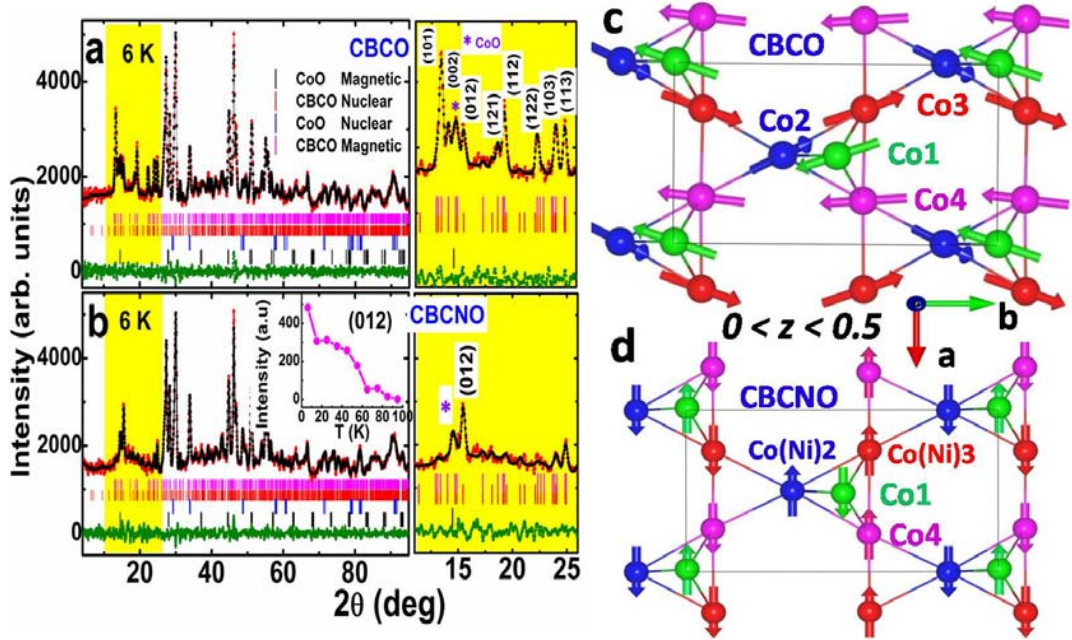

**FIG.3.** (Color online). Observed and calculated diffractions patterns and their difference at 6 K for CBCO (a) and CBCNO (b) samples. Red circles are experimental data points and black curve is calculated pattern. Green line at the bottom shows the difference between the measured and calculated patterns. Inset of (b) shows the temperature variation of (012) peak intensity. (c) and (d) are spin structures of CBCO and CBCNO samples at 6 K view along the [001] direction in the range $0 < z < 0.5$. The different Co sites are labelled with same scheme as in Ref. [15].

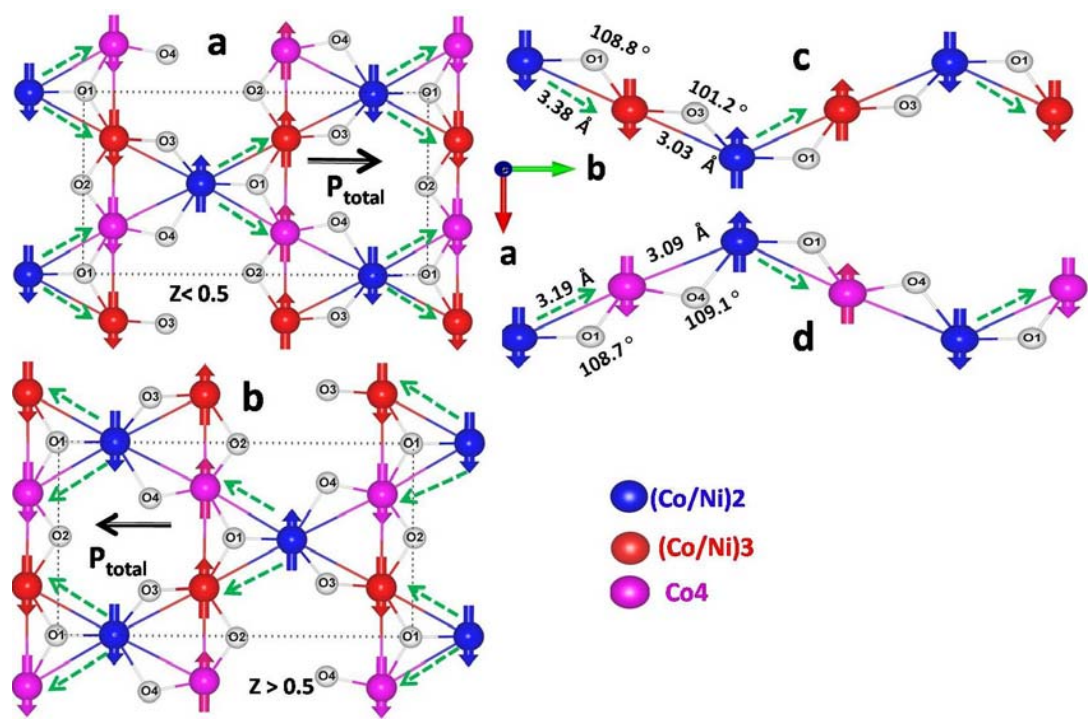

**FIG.4.** (Color online). (a) and (b) Pictorial reperesentation of the charge and magnetic ordering mechanism of the appearence of polarization (green dashed arrows) in kagome layers for CBCNO sample. (c,d) Co2-Co3 and Co3-Co4 zig zag chains along b-direction in kagome layer ( z < 0.5).

Supplementary Information

**Sample preparation and experimental techniques**

The samples of $CaBaCo_{4-x}Ni_xO_7$ (x=0, 0.05 and 0.10) were prepared by standard solid state reaction method, using high purity $CaCo_3$, $BaCo_3$, $NiO$ and $Co_3O_4$. Stoichiometric mixtures of these powders were ground thoroughly, heated in air at 900 °C for 12 h, 1100 °C for 24 h and quenched to room temperature. Phase purity of the samples is confirmed initially by x-ray powder diffraction at room temperature where all samples are assigned to orthorhombic structure with *Pbn2₁* space group. The neutron diffraction measurements on x=0 and 0.10 samples were carried out on the PD2 powder neutron diffractometer ($\lambda$=1.2443 Å) at Dhruva reactor, Bhabha Atomic Research Centre, Mumbai, India. The powder samples of approximately 5g were packed in a cylindrical vanadium container and attached to the cold finger of a closed cycle helium refrigerator. Rietveld refinement of the neutron diffraction patterns were carried out using the FULLPROF program [4]. Magnetic measurements were carried in a commercial VSM SQUID magnetometer (Quantum Design, USA).Dielectric and pyroelectric current measurements were done in parallel plate capacitor geometry using Agilient 4291A and Keithley electrometer 6517A respectively. For dielectric and pyroeletric current ($I_p$) measurements pellets of thickness 0.5 mm and diameter of 5 mm were used.

**Pyroelectric behaviour of CBCO:**

Pyroelectric and ferroelectric materials have the same prerequisite symmetry and ferroelectric materials form a subset of pyroelectric materials, in which spontaneous electric polarization can be reversed by an external electric field. Spontaneous polarization of the pyroelectric materials is strongly temperature dependent due to their crystallographic structure. A pyroelectric material generates a voltage (current) when they are heated or cooled, through transition temperature even without electric field. The measured $I_p$ in heating and cooling paths for different heating/cooling rates, in the absence of poling field is shown in the Fig.S1 (a). It can be noticed from the Fig.S1 that (a) the sign of the $I_p$ is opposite for cooling and heating paths; this behavior is typical for pyroelectric materials as depicted in Fig S1 (b); because $I_p \propto dT/dt$ (T: temperature, t: time). Pyroelectric materials possess permanent dipoles and thus induce non-zero net polarization (Ps) and hence attract free charges to its surface. If there is no change in temperature, there is no change in Ps and no current flows through the external circuit. When cooling (heating) the sample across the transition temperature the dipolar ordering attract the ambient charges (randomization of dipoles release charge) to the surface of the electrodes resulting a flow of current (in opposite) direction in the circuit.

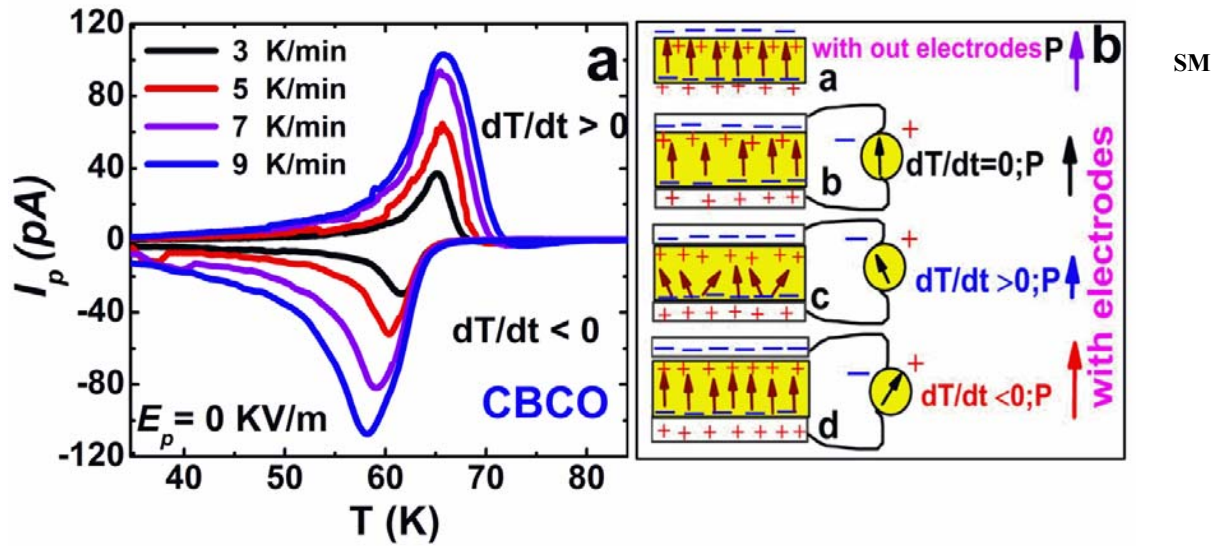

**FIG.S1** (a) Heating and cooling rate dependence of pyrocurrent ($I_p$) for CBCO measured in absence of poling field. (b) Schematic model for the observed phenomenon in (a).

**Ferroelectric behaviour of CBCNO (Ni 5%) samples:**

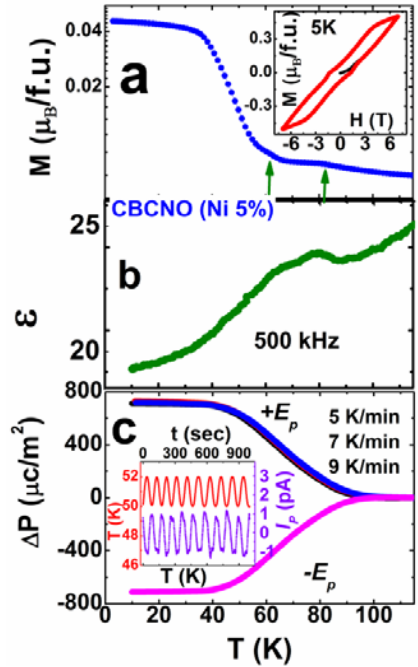

**SM FIG.S2.** Temperature dependent (a) magnetization (logarithmic scale) (b) dielectric constant and (c) $\Delta P$ of CBCNO (Ni 5%). Inset (a) shows the M-H at 5K and inset (c) shows the variation of the pyrocurrent and temperature with time in the temperature interval of 50-52 K.

Temperature dependence of magnetization for 5% Ni substituted samples is shown in Fig.S2 (a); two magnetic transitions at 60 K and 82 K can be noticed similar to 10% Ni sample in Fig.1 (b). The temperature dependent dielectric permittivity of the 5% Ni substituted samples is shown in the Fig.S2 (b). The 5% Ni substituted samples shows two peaks at 60 K and ~82 K, which is also similar to the 10% Ni doped sample (Fig.2(c)). The switching of the polarization for the opposite poling fields and same magnitude of the polarization for various ramp rates suggests the ferroelectric behaviour of the 5% Ni sample. The intrinsic ferroelectric switching behavior in 5 % Ni doped CBCO sample is confirmed by time vs temperature cycle measurements between the 50-52 K temperature intervals (done after waiting time of 40 minutes at 50 K).

**The neutron diffraction data on the CBCO and 10% Ni doped CBCO is given below.**

TABLE Ia. Atomic parameters for CBCO at 6 K.

Space group: $Pbn2_1$. Cell parameters: a = 6.2556(5) (Å), b = 11.0383(9) (Å), c = 10.1563(8) (Å). V= 701.3(1) (Å)$^3$

| Atom | x | y | z | B | reliability factors |
|---|---|---|---|---|---|
| Ca | -0.005 (4) | 0.674 (2) | 0.873 (2) | 0.3(1) | Rp (%) = 2.77 |
| Ba | 0.003 (2) | 0.667 (2) | 0.50000 | 0.3(1) | |
| Co1 | 0.008 (5) | -0.005 (4) | 0.941 (3) | 0.2(4) | Rwp (%) = 3.53 |
| Co2 | 0.005 (5) | 0.161 (3) | 0.701 (4) | 0.2(4) | |
| Co3 | 0.248 (5) | 0.092 (3) | 0.196 (4) | 0.2(4) | $\chi^2$ (%) = 2.33 |
| Co4 | 0.261 (5) | 0.914 (2) | 0.677 (3) | 0.2(4) | |
| O1 | -0.012 (4) | -0.005 (3) | 0.255 (3) | 0.1(2) | |
| O2 | -0.012 (4) | 0.496 (2) | 0.231 (3) | 1.2 (3) | |
| O3 | 0.777 (5) | 0.258 (2) | 0.789 (3) | 0.9 (4) | |
| O4 | 0.728 (4) | 0.748 (2) | 0.216 (3) | 0.5 (3) | |
| O5 | -0.047 (2) | 0.157 (1) | 0.501 (3) | 1.0 (3) | |
| O6 | 0.206 (2) | 0.108 (1) | -0.002 (4) | 0.2(2) | |
| O7 | 0.266 (2) | 0.947 (1) | 0.499 (4) | 0.5 (2) | |
| Co site magnetic moments at 6K ||||||
| | Mx | My | Mz | M total | |
| Co1 | -1.4 (1) | -2.2 (1) | 0 | 2.7(1) | |
| Co2 | 1.3 (1) | 1.4 (2) | 1.4 (2) | 2.2(2) | |
| Co3 | 1.7 (2) | 2.8 (1) | 0.9 (2) | 3.4(2) | |
| Co4 | -0.2(2) | -2.0(1) | 0 | 2.0(2) | |

**TABLE Ib.** Atomic parameters for CBCO at 300 K. Space group: $Pbn2_1$. Cell parameters: a = 6.2837(6) (Å), b = 11.0032(9) (Å), c = 10.1857(8) (Å). V= 704.2 (1) (Å)$^3$

| Atom | x | y | z | B | reliability factors |
|------|---|---|---|---|---------------------|
| Ca | 0.003 (3) | 0.676 (2) | 0.872 (2) | 0.4(1) | Rp (%) = 2.65 |
| Ba | -0.002 (2) | 0.666 (2) | 0.50000 | 0.4(1) | |
| Co1 | -0.001 (4) | -0.007 (3) | 0.947 (2) | -0.4(1) | Rwp (%) = 3.31 |
| Co2 | 0.007 (5) | 0.177 (2) | 0.698 (2) | -0.4(1) | |
| Co3 | 0.225 (4) | 0.100 (2) | 0.202 (2) | -0.4(1) | χ2 (%) = 2.18 |
| Co4 | 0.242 (5) | 0.908 (2) | 0.677 (2) | -0.4(1) | |
| O1 | -0.001 (4) | 0.001 (2) | 0.263 (2) | 0.6(2) | |
| O2 | -0.005 (3) | 0.493 (2) | 0.240 (2) | 1.0(2) | |
| O3 | 0.787 (2) | 0.261 (1) | 0.791 (2) | 0.4(2) | |
| O4 | 0.746 (2) | 0.746 (1) | 0.219 (2) | 1.0(2) | |
| O5 | -0.052 (2) | 0.161 (1) | 0.502 (2) | 0.9(2) | |
| O6 | 0.213 (2) | 0.1020 (9) | -0.000 (3) | 0.7(2) | |
| O7 | 0.262 (2) | 0.944 (1) | 0.503 (3) | 0.8(2) | |

**TABLE IIa.** Atomic parameters for CBCNO (Ni 10%) at 6 K. Space group: $Pbn2_1$. Cell parameters: a = 6.2561 (6) (Å), b = 11.0246 (9) (Å), c = 10.1673 (9) (Å). V= 701.243(0.104) (Å)$^3$

| Atom | x | y | z | B | Site Occupancy |
|------|---|---|---|---|----------------|
| Ca | -0.007 (2) | 0.676 (1) | 0.877 (1) | 0.05(8) | 1.0 |
| Ba | 0.003 (1) | 0.663 (1) | 0.50000 | 0.05(8) | 1.0 |
| Co1 | 0.019 (3) | 0.000 (3) | 0.947 (2) | -0.16(10) | 1.0 |
| Co2 | 0.013(3) | 0.178 (1) | 0.690 (2) | -0.16(10) | 0.96(1) |
| Ni2 | | | | | 0.04(1) |
| Co3 | 0.250 (3) | 0.094 (2) | 0.200 (2) | -0.16(10) | 0.95(1) |
| Ni3 | | | | | 0.05(1) |
| Co4 | 0.258 (4) | 0.924 (1) | 0.700 (2) | -0.16(10) | 1.0 |
| O1 | -0.019 (2) | 0.007 (1) | 0.258 (1) | 0.59(16) | 1.0 |
| O2 | -0.004 (2) | 0.495 (1) | 0.235 (1) | 0.58(15) | 1.0 |
| O3 | 0.778 (2) | 0.2603 (9) | 0.794 (1) | 0.42(15) | 1.0 |
| O4 | 0.717 (2) | 0.7431 (9) | 0.221 (1) | 0.34(16) | 1.0 |
| O5 | -0.050 (2) | 0.162 (1) | 0.511 (2) | 0.96(17) | 1.0 |
| O6 | 0.207 (2) | 0.1136 (8) | -0.002 (2) | 1.13(17) | 1.0 |
| O7 | 0.266 (2) | 0.9456 (7) | 0.512 (2) | 0.86(15) | 1.0 |
| Co site magnetic moments at 6K | | | | | |
| | Mx | My | Mz | M total | reliability factors |
| Co1 | -2.03(4) | 0 | 0 | -2.03(4) | Rp (%) = 1.77 |
| Co2 Ni2 | 1.84(8) | 0 | 0 | 1.84(8) | Rwp (%) = 2.24 |
| Co3 Ni3 | 1.1(3) | 0 | 0 | 1.1(3) | χ2 (%) = 2.08 |
| Co4 | 1.4(2) | 0 | 0 | 1.4(2) | |

**TABLE IIb.** Atomic parameters for CBCNO (Ni 10%) at 300 K. Space group: $Pbn2_1$. Cell parameters: a = 6.2832(3) (Å), b = 10.9981(5) (Å), c = 10.1817(4) (Å). V= 703.589 (0.056) (Å)$^3$

| Atom | x | y | z | B | Site Occupancy |
|---|---|---|---|---|---|
| Ca | 0.002 (2) | 0.666 (1) | 0.8754 (8) | 0.41(6) | 1.0 |
| Ba | 0.004 (1) | 0.6573 (8) | 0.50000 | 0.41(6) | 1.0 |
| Co1 | 0.014 (3) | -0.002 (2) | 0.948 (1) | -0.13(7) | 1.0 |
| Co2 | 0.0238 (2) | 0.185(1) | 0.691 (2) | -0.13(7) | 0.96(1) |
| Ni2 | | | | | 0.04(1) |
| Co3 | 0.242 (3) | 0.091(1) | 0.195 (1) | -0.13(7) | 0.95(1) |
| Ni3 | | | | | 0.05(1) |
| Co4 | 0.248 (3) | 0.926 (1) | 0.696 (1) | -0.13(7) | 1.0 |
| O1 | -0.005 (3) | -0.005 (1) | 0.260 (1) | 0.87(10) | 1.0 |
| O2 | -0.003 (2) | 0.4919 (93) | 0.238 (1) | 1.21(14) | 1.0 |
| O3 | 0.786 (1) | 0.2566 (90) | 0.789 (1) | 0.84(13) | 1.0 |
| O4 | 0.739 (2) | 0.7492 (83) | 0.222 (1) | 0.95(14) | 1.0 |
| O5 | -0.060 (1) | 0.1595 (87) | 0.505 (1) | 1.96(15) | 1.0 |
| O6 | 0.210 (1) | 0.1022 (68) | -0.002 (2) | 1.78(16) | 1.0 |
| O7 | 0.262 (1) | 0.9415 (56) | 0.513 (1) | 0.77(10) | 1.0 |
| reliability factors Rp (%) = 1.24 Rwp (%) = 1.61 χ2 (%) = 6.93 | | | | | |

**TABLE III.** Comparison of Interatomic distances at 300K and 6K for CBCO and CBCNO (Ni 10 %) samples.

| BOND | CBCO | | CBCNO | |
|---|---|---|---|---|
| | 300K | 6K | 300K | 6K |
| Ca-O2 | 2.30(2) | 2.37(3) | 2.232(15) | 2.375(18) |
| O3 | 2.21(2) | 2.12(4) | 2.241(15) | 2.112(18) |
| O4 | 2.34(2) | 2.33(4) | 2.352(15) | 2.290(18) |
| O5 | 2.25(2) | 2.30(3) | 2.354(16) | 2.275(19) |
| O6 | 2.35(3) | 2.38(3) | 2.310(16) | 2.348(18) |
| O7 | 2.41(3) | 2.34(3) | 2.375(14) | 2.394(19) |
| | | | | |
| Ba-O2 | 3.26(2) | 3.32(3) | 3.228(12) | 3.268(14) |
| O2 | 3.01(2) | 2.96(3) | 2.927(12) | 2.952(14) |
| O3 | 3.615(20) | 3.56(3) | 3.633(12) | 3.635(15) |
| O3 | 2.649(20) | 2.68(3) | 2.690(12) | 2.640(15) |
| O4 | 3.391(19) | 3.47(3) | 3.439(12) | 3.470(14) |
| O4 | 2.884(19) | 2.77(3) | 2.886(12) | 2.815(14) |
| O5 | 3.482(18) | 3.405(18) | 3.490(12) | 3.423(14) |
| O5 | 2.803(18) | 2.855(18) | 2.795(12) | 2.837(14) |
| O6 | 3.449(19) | 3.56(2) | 3.399(12) | 3.561(14) |
| O6 | 2.878(20) | 2.80(2) | 2.970(12) | 2.797(14) |
| O7 | 3.48(2) | 3.50(2) | 3.523(11) | 3.529(14) |
| O7 | 2.869(20) | 2.82(2) | 2.794(11) | 2.799(14) |
| | | | | |
| Co1-O1 | 1.88(3) | 1.89(4) | 1.917(16) | 1.92(2) |
| O5 | 1.82(4) | 1.80(5) | 1.85(2) | 1.91(3) |
| O6 | 1.88(3) | 1.85(4) | 1.76(2) | 1.80(3) |
| O7 | 1.87(3) | 1.92(4) | 1.97(2) | 1.99(2) |
| Co2-O1 | 2.07(3) | 1.81(5) | 2.100(18) | 2.158(20) |
| O3 | 1.91(3) | 1.99(4) | 1.967(17) | 2.02(2) |
| O4 | 1.78(3) | 1.95(4) | 1.686(16) | 1.93(2) |
| O5 | 2.04(3) | 2.06(5) | 1.98(2) | 1.87(3) |
| Co3-O1 | 1.89(3) | 2.04(4) | 1.99(2) | 2.02(2) |
| O2 | 2.15(3) | 1.99(4) | 2.01(2) | 1.96(2) |
| O3 | 1.82(3) | 1.91(4) | 1.958(17) | 1.88(2) |
| O6 | 2.06(4) | 2.04(6) | 2.01(2) | 2.08(3) |
| Co4-O1 | 2.01(4) | 2.02(4) | 1.87(2) | 1.78(3) |
| O2 | 2.03(3) | 1.82(4) | 1.86(2) | 1.77(3) |
| O4 | 1.75(3) | 1.84(3) | 1.945(16) | 1.871(19) |
| O7 | 1.82(4) | 1.85(5) | 1.875(20) | 1.92(3) |